# SpectraSensML Software: Mastering Complete Spectral Information for Luminescence Thermometry 2.0

*Aleksandar Ćirić*, Zoran Ristić, Tamara Gavrilović, Anđela Rajčić, Snežana Đurković, Željka Antić, and Miroslav D. Dramićanin**

Centre of Excellence for Photoconversion, Vinča Institute of Nuclear Sciences - National Institute of the Republic of Serbia, University of Belgrade, P.O. Box 522, Belgrade 11251, Serbia

*Corresponding authors email: aleksandar.ciric@ff.bg.ac.rs, dramican@vinca.rs

## Abstract

Luminescence thermometry has evolved through decades of research focused on optimising materials and on extracting temperature information from isolated spectral features such as luminescence intensity ratios, bandwidth, line shift and excited-state lifetime. Despite extensive material development, these conventional methods remain fundamentally limited by construction: only a small subset of pre-selected spectral features is exploited, while the bulk of the temperature-relevant information encoded in the full spectrum is systematically discarded. A paradigm shift is presented here, Luminescence Thermometry 2.0 (LT 2.0), implemented through the newly developed SpectraSensML platform, in which machine learning regression operates on the entire emission spectrum to deliver temperature readout. The approach is demonstrated on a $Yb^{3+}$-doped phosphor emitting in the near-infrared biological transparency window across 125 to 700 K. $Yb^{3+}$ is a particularly demanding case: only the single $^2F_{5/2}$ multiplet emits, and its weakly thermally coupled Stark sub-levels yield modest sensitivity under conventional intensity-ratio thermometry. Nineteen regression algorithms drawn from four families, namely tree ensembles, physics-aware regression models, kernel and instance methods, and neural networks, are systematically benchmarked. A sensor-fusion estimator that combines the first three principal components reaches an average root-mean-square error of 0.36 K on an unseen-temperature test set, a seven-fold improvement over the best luminescence intensity ratio variant. Standard normal variate (SNV) normalisation is identified as the most effective preprocessing strategy because it isolates the band-shape deformations that encode temperature. Single-component approaches that rely on the first principal component alone are shown to be quantitatively sub-optimal: multi-component regressors that exploit the first three principal components reduce the temperature uncertainty by close to an order of magnitude. The structural reason behind the failure of decision-tree ensembles on unseen temperatures is explained: their piecewise-constant predictions cannot interpolate beyond training set-points. The open-source SpectraSensML application used to obtain the results is released alongside the manuscript to enable reproducible community benchmarks.

## 1. Introduction

### 1.1 Luminescence Thermometry: Evolution of the Field

Luminescence thermometry exploits temperature-induced changes in the optical response of luminescent materials to provide remote, contactless temperature readout [1]. Over the past two decades the field has expanded rapidly, driven by diverse applications ranging from biomedical imaging and diagnostics to microelectronics thermal management, catalytic reactor monitoring and nanoscale heat-transport studies [2–4].

The methodology has evolved through successive generations, each marked by progressively greater sophistication in both material selection and data processing. Current state-of-the-art techniques can be classified by the spectral feature being exploited. Traditional feature-based methods (LT 1.0) rely on extracting and calibrating isolated spectral parameters against temperature. The most widely used steady-state techniques are (i) the luminescence intensity ratio (LIR), which most often exploits the Boltzmann distribution between two thermally coupled electronic levels; (ii) bandwidth thermometry, which measures temperature-dependent line broadening induced by electron–phonon coupling; and (iii) line-shift thermometry, which tracks the spectral position of emission bands as they shift with temperature through crystal-field variations and thermal expansion [5]. Time-resolved methods based on luminescence lifetime are also widely used but require pulsed excitation and time-gated detection.

The LIR paradigm has dominated the field for decades. In this approach the temperature is recovered from the ratio of their integrated emission intensities. For the thermally coupled energy levels, LIR temperature dependence takes the form [6]:

$$LIR(T) = \frac{I_H}{I_L} = B \cdot \exp\left(-\frac{\Delta E}{k_B\, T}\right) \tag{1}$$

where $I_H$ and $I_L$ denote the integrated intensities of the energetically higher (H) and lower (L) emission levels, respectively, $\Delta E$ is the energy gap between the two thermally coupled levels, B is a temperature-invariant prefactor that absorbs degeneracies and radiative-rate ratios, and $k_B$ is the Boltzmann constant. The relative thermal sensitivity follows directly:

$$S_r = \frac{1}{LIR}\left|\frac{d\, LIR}{dT}\right| = \frac{\Delta E}{k_B T^2} \tag{2}$$

meaning that larger energy gaps yield higher sensitivity. The most studied Boltzmann thermometers are based on the green emissions of $Er^{3+}$ ($^2H_{11/2}$ and $^4S_{3/2} \rightarrow {}^4I_{15/2}$, $\Delta E \approx 700$ cm$^{-1}$) and the blue emissions of $Dy^{3+}$ ($^4I_{15/2}$ and $^4F_{9/2} \rightarrow {}^6H_{15/2}$, $\Delta E \approx 1125$ cm$^{-1}$), owing to their nearly ideal energy separations and strong, well-separated emission bands [7].

Enhanced single-feature methods (LT 1.1) include innovations such as single-band ratiometric thermometry [8,9], the luminescence intensity ratio squared [10], multilevel cascade LIR [11], and deconvolution-assisted methods [12]. These approaches aim to boost sensitivity while preserving the conceptual simplicity of a single-number readout. Statistical and regression-based approaches (LT 1.5) introduce multivariate techniques such as multiparameter linear regression [13,14], and sensor fusion by inverse-variance weighting to combine several temperature-dependent features [15]. These methods still rely on manual feature extraction and selection, inherently limiting the use of the full information content of luminescence spectra.

All conventional approaches share a fundamental limitation: only a small subset of spectral features is extracted, typically one or two narrow integration windows, while the vast majority of temperature-relevant information encoded in the full spectrum is systematically discarded [16]. The LIR by construction reduces an entire spectrum to a single scalar feature, ignoring band shape, the relative intensities of sub-bands, temperature-dependent broadening, background curvature and other subtle yet information-rich spectral changes. Even after years of optimisation, these methods reach performance plateaus that cannot be overcome without a methodological change. This is the research gap addressed by the LT 2.0 paradigm (see Figure 1).

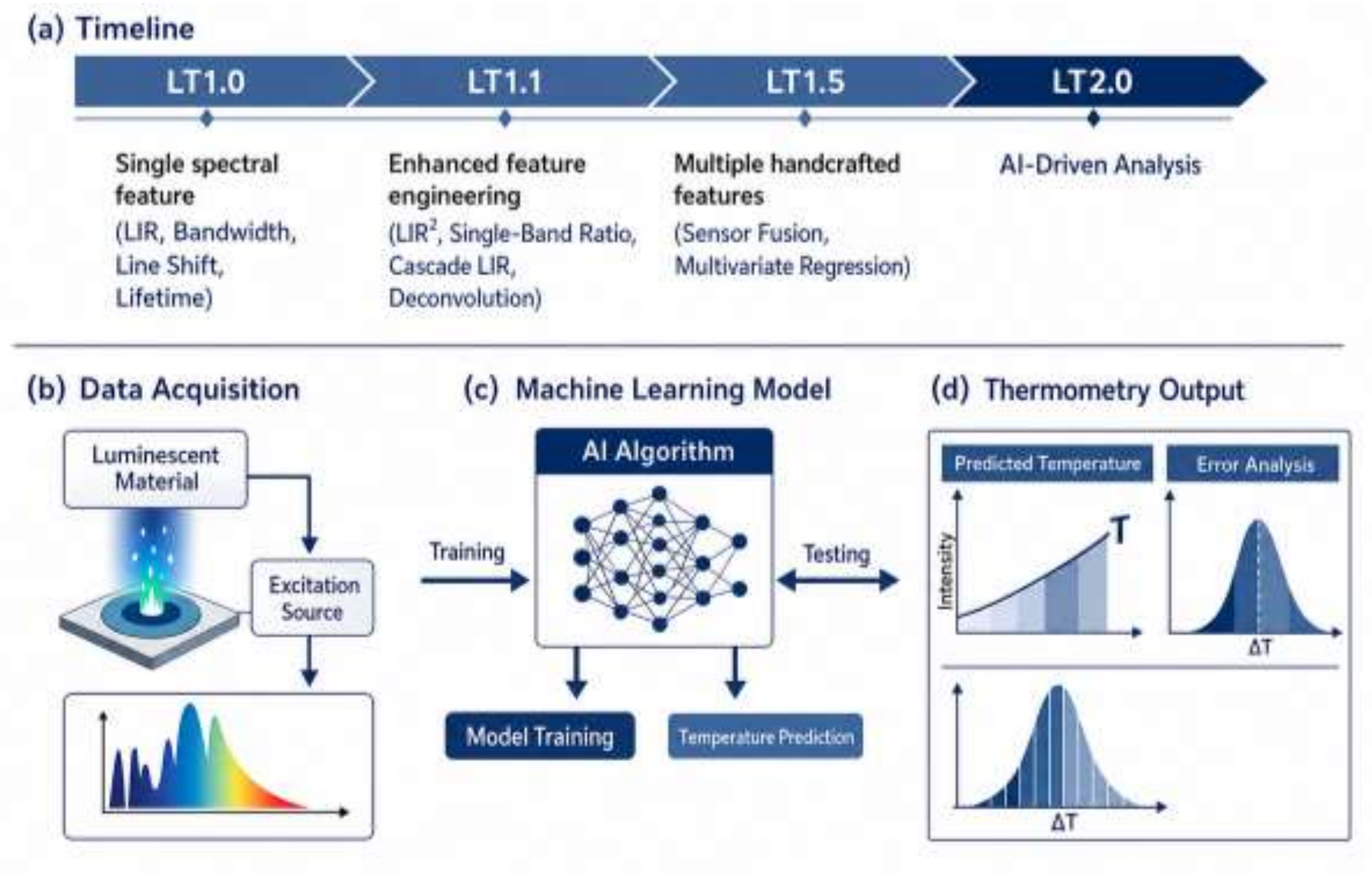

**Figure 1. Schematic overview of the Luminescence Thermometry 2.0 framework. (a) Evolution of luminescence thermometry toward full-spectrum machine learning. (b) Spectral acquisition from the luminescent material under optical excitation. (c) Data-driven workflow for preprocessing, machine-learning model training, calibration, and temperature prediction. (d) Thermometric output with predicted temperature and statistical error analysis.**

## 1.2 The NIR and $Yb^{3+}$ Case Study

The near-infrared spectral range is of particular importance for biomedical applications: tissue absorbs and scatters near-infrared radiation much less than visible light, defining the so-called biological transparency windows [17]. Among trivalent lanthanide ions, $Yb^{3+}$ is the workhorse near-infrared emitter, widely used as an activator in solid-state laser systems and as an upconversion sensitiser, owing to its large absorption cross-section [18]. The single inter-multiplet transition $^2F_{5/2} \rightarrow {}^2F_{7/2}$ produces a broad, structured, NIR emission band. Sensitisation through UV-absorbing co-dopants and host charge-transfer states, as in $YVO_4$:$Yb^{3+}$, can produce intense near-infrared emission suitable for fibre-optic sensing, solar-cell diagnostics and microfluidic temperature monitoring [19].

$Yb^{3+}$ is, however, a particularly demanding case for conventional LIR thermometry. The ion possesses only one emitting multiplet, $^2F_{5/2}$, with no readily excited higher electronic level within the spectral domain of practical luminescence. The only available thermally coupled states are the Stark sub-levels of $^2F_{5/2}$ split by the crystal field, with energy separations typically in the range of several hundred $cm^{-1}$. Zhang et al. demonstrated that LIR-based $Yb^{3+}$ thermometers can achieve optimal performance only at 200 K to 400 K range, and with limited sensitivities [20].

More fundamentally, the LIR approach applied to $Yb^{3+}$ emission uses at most two narrow spectral windows of an otherwise information-rich band [21]. Luminescence spectra of lanthanide ions are, however, intrinsically multi-dimensional: every Stark transition contributes its own temperature dependence through population redistribution, homogeneous and inhomogeneous broadening, electron–phonon coupling and crystal-field modulation. Every feature carries a piece of the temperature information. The natural question is therefore whether a learning algorithm operating on the full spectrum, encompassing all Stark transitions simultaneously, can recover the hidden information that LIR systematically discards. This challenging spectra on a well-known material were thus selected for testing the LT 2.0 methods and the software presented in this paper.

### 1.3 The Paradigm Shift: Introducing Luminescence Thermometry 2.0 (LT 2.0)

Luminescence Thermometry 2.0 represents not merely an incremental improvement but a fundamental shift that addresses the persistent research gap in the field. The core idea is straightforward yet transformative: instead of manually selecting one or a few spectral features, the entire spectrum is treated as the temperature indicator and a machine-learning regression algorithm is used to automatically learn the spectrum-to-temperature mapping. By exploiting complete spectral information rather than selected features, LT 2.0 accesses all temperature-relevant information that conventional methods discard.

Prior work has demonstrated PCA-based thermometry using only the first principal component (PC1) [22–24]. While that approach showed improvement over traditional LIR, it inherits a subtle limitation: by discarding PC2, PC3 and higher components it ignores complementary information encoded in spectral asymmetries, fine structure and secondary temperature-dependent features. Multi-component models that exploit multiple principal components are hypothesized to consistently outperform PC1-only regressors. Complementary to other methods, the sensor-fusion approach, in which individual PC-based predictors are combined with inverse-variance weighting, provides a metrology-inspired framework for extracting and integrating information from multiple principal components.

Recent years have seen the rapid adoption of machine learning for spectroscopic regression, with feed-forward neural networks [25–29], one-dimensional convolutional neural networks applied to raw spectra [30], random-forest regressors [31], and gradient-boosted ensembles reported in the literature [32]. Most such studies, however, focus on a single algorithm and a single dataset, and rarely include a fair comparison under identical experimental conditions. Although spectral outputs from physical sensors depend continuously on the calibrated quantity, the use of machine learning to model this continuous behaviour has not been systematically investigated. A reproducible benchmark of nineteen regression algorithms applied to a $Yb^{3+}$ near-infrared luminescence dataset spanning 125 K to 700 K is presented here, together with the open-source application SpectraSensML used to obtain the results. The most accurate model families for continuous-quantity sensing are identified, the benefit of full-spectrum machine learning over LIR is quantified, preprocessing best practice is established, and the mechanistic reason behind the failure of decision-tree ensembles to generalise across temperature is explained.

## 2. Experimental

The investigated material is yttrium orthovanadate doped with 4 mol% $Yb^{3+}$ ($YVO_4$:$Yb^{3+}$). Upon ultraviolet excitation at 265 nm the phosphor exhibits intense emission centred near 980 nm corresponding to the $^2F_{5/2} \rightarrow {}^2F_{7/2}$ transition. The $Yb^{3+}$ concentration of 4 mol % balances emission intensity with concentration-quenching limits.[19] Full synthesis procedure, batch-to-batch reproducibility, XRD, FE-SEM morphology and UV-Vis-NIR diffuse reflectance characterization are reported in sections S1-3 in the Supplementary Information.

Temperature-dependent emission spectra were recorded with the sample mounted in a calibrated MicroOptik liquid-nitrogen-cooled heating and cooling stage (temperature stability ±0.1 K). Excitation was provided by a 265 nm UV LED module (Ocean Insight). The emission was collected through a quartz fibre bundle and analysed with an Ocean Insight NIRQuest spectrometer covering 897 to 1708 nm with 512 pixels (spectral resolution 1.6 nm), under an integration time of 1500 ms.

The dataset comprises 80 temperature set-points spanning 125 to 700 K with 100 spectra acquired at each set-point, for a total of approximately 8000 spectra. Three disjoint splits were used. Set-points whose nominal Kelvin value ends in 0, for example 125, 110, …, 700 K, were assigned to a train/validation pool, while the remaining 22 set-points (123, 148, 173, 198, 223, 248, 323, 348, 373, 398, 423, 448, 473, 498, 523, 548, 573, 598, 623, 648, 673 and 698 K) were assigned to a unseen test split (*test_unseen*) that is fully withheld from training. Within each train set-point, 80% of spectra (randomly selected) were placed into the training set and the remaining 20 % into an internal validation set. The *test_unseen* split therefore serves as the strictest possible generalisation test because no spectrum at these set-points is present at training time.

All spectra were cropped to the $Yb^{3+}$ emission band (900 nm to 1100 nm, 124 pixels), the per-spectrum minimum was subtracted as a baseline, and the resulting spectra were normalised. Three normalisation strategies were considered, MAX, AREA and SNV (see Section 3.2), and SNV was retained as the default because it consistently yielded the lowest test errors across all model families. Principal component analysis was fitted on the training set only and applied to the validation and *test_unseen* splits. Three components were retained as default. Implementation details and per-model hyperparameters are listed in Supplementary Information.

Where a model exposed tunable hyperparameters, these were selected by minimizing the *validation* set RMSE. Models with a single natural configuration were used with fixed settings, and the univariate PC1 polynomial order was chosen by the Bayesian information criterion. Performance was reported separately on *training*, *validation* and *test_unseen* sets using sensor-style metrics: root-mean-square error (RMSE), mean absolute error (MAE), maximum absolute error (MaxErr), bias, and the temperature resolution $\delta T$ (the standard deviation of residuals). Models were ranked by the arithmetic mean of *validation* and *test_unseen* RMSE so that both interpolation precision and extrapolation behaviour contributed equally to the ranking.

All preprocessing, dimensionality reduction, model training, evaluation and statistical analyses were carried out with the here-created open-source SpectraSensML thermometry application (GPLv3 license), described in Section 4.1.

## 3. Theoretical Framework

### 3.1 From Boltzmann Thermometry to Machine Learning

The theoretical foundation of conventional LIR thermometry rests on the Boltzmann distribution (Eq. 1). Eq. 2 reveals the fundamental trade-off in LIR thermometry: larger energy gaps yield higher sensitivity, but the thermal coupling weakens (the population of the upper level becomes negligible), reducing the signal-to-noise (SNR) ratio. For $Yb^{3+}$ the small Stark splitting impose a hard ceiling on the achievable sensitivity.

Machine-learning approaches abandon the Boltzmann ansatz entirely. Instead of assuming a specific parametric form, a supervised regression algorithm learns a mapping $f_\theta : S(\lambda) \rightarrow T$ from training data, where $S(\lambda)$ is the emission spectrum (a vector of intensities at wavelengths $\lambda_1, \lambda_2, \ldots, \lambda_n$), $\theta$ collects the model parameters, and T is the temperature. The model prediction for the $i^{th}$ spectrum is denoted $\hat{T}_i \equiv f_\theta(S_i)$. For parametric regressors such as polynomial regression, kernel ridge regression and multilayer perceptrons, $\theta$ is obtained by minimising a regularised mean-squared-error objective [33]:

$$L(\theta) = \frac{1}{N} \sum_i \left( \hat{T}_i - T_i \right)^2 + \lambda \, \Omega(\theta) \tag{3}$$

where the sum runs over $N$ training spectra, $\lambda \geq 0$ controls the strength of the regularisation and $\Omega(\theta)$ is a model-specific penalty (weight decay for the multilayer perceptron, ridge penalty for kernel ridge regression, mixed penalties for ElasticNet, and so on). Not every regressor benchmarked here is fitted by minimising Eq. 3: tree ensembles minimise a local impurity criterion at each split, Gaussian-process regression maximises a marginal log-likelihood, and k-nearest neighbours has no training objective at all. Importantly, $f_\theta$ is not constrained to take a Boltzmann form, it can be a polynomial, spline, kernel expansion, neural network or any other flexible functional form. This flexibility allows the model to capture the non-linear temperature dependences arising from multiple overlapping Stark transitions, intensity redistribution, broadening and baseline drift, phenomena that a two-band LIR cannot represent.

Directly regressing temperature on several hundred strongly correlated spectral channels is, however, ill-conditioned and prone to overfitting on modest datasets. In practice, the spectrum is therefore first compressed by principal component analysis (PCA, see Section 3.3) into a compact, uncorrelated score vector, and it is this vector, rather than the raw spectrum, that serves as the input to $f_\theta$ for most of the regressors benchmarked here.

Once a model is fitter, by whichever training criterion is appropriate to its family, its sensor-style performance is quantified on an independent evaluation split by the root-mean-square error of its temperature predictions, expressed in the same physical units as the measured quantity:

$$\mathrm{RMSE} = \sqrt{\frac{1}{N}\sum_{i}\left(\hat{T}_i - T_i\right)^2} \quad (4)$$

with $N$ the number of evaluated spectra (training, validation or unseen-test split), $T_i$ the set-point reference temperature of the $i^{\text{th}}$ spectrum and $\hat{T}_i$ the corresponding model prediction. For the parametric regressors fitted by Eq. 3, the RMSE evaluated on the training set coincides with the square root of that objective's data-fit term with the regularization penalty removed, while for all other models it is purely an evaluation metric.

In sensor-metrology language, the RMSE bundles together the accuracy (closeness of the predicted mean to the true value, which is dominated by the bias of the calibration) and the precision (spread of individual predictions around their mean, temperature resolution) of a thermometer. When evaluated on a split that contains many spectra per set-point, $\mathrm{RMSE}^2$ approximately equals $\mathrm{bias}^2 + \delta T^2$ [34], with the bias defined as the mean residual and $\delta T$ defined as the standard deviation of residuals (see Figure 2). The RMSE constitutes the primary figure of merit throughout this work.

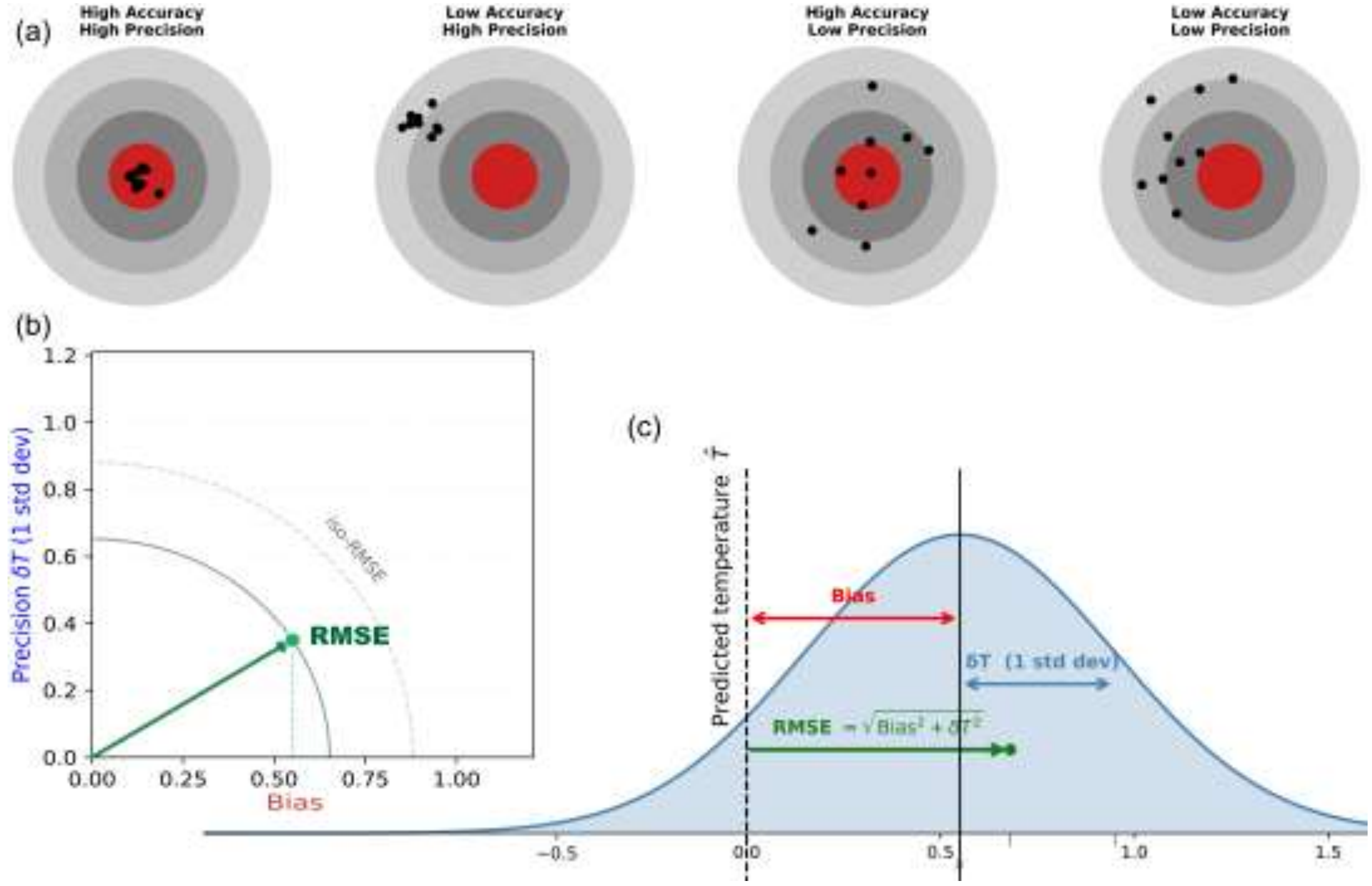

**Figure 2. (a) Precision, accuracy, and RMSE illustrated by analogy with target diagrams. (b) 2D representation showing how bias and temperature resolution combine into RMSE, and (c) demonstrating those relations by probability read-out on temperature scale.**

An additional metric of practical interest is the mean absolute error (MAE), which reports the average magnitude of the residuals without squaring. MAE is less sensitive than RMSE to outliers and individual large residuals, and is often preferred for reporting typical sensor errors. It is defined as [35]:

$$\text{MAE} = \frac{1}{N}\sum_{i}\left|\hat{T}_i - T_i\right| \tag{5}$$

and is reported alongside RMSE throughout the manuscript. The maximum absolute error, MaxErr = $\max_i |\hat{T}_i - T_i|$, captures the worst-case behaviour and is the most relevant figure of merit for safety-critical or specification-driven deployments. Other relevant metrics are also given in the Supplementary Information.

## 3.2 Data Preprocessing and Normalisation

Robust preprocessing is essential for extracting temperature information from raw luminescence spectra, which are subject to instrumental artefacts, excitation-power fluctuations, detector dark current, stray light and environmental noise [36,37]. The preprocessing pipeline employed here consists of background subtraction, cropping to the active emission band and per-spectrum normalisation.

Three normalisation strategies are commonly used in vibrational and luminescence spectroscopy:

1) MAX normalisation divides each spectrum by its maximum intensity, anchoring all spectra to a common peak height. The procedure is vulnerable to detector saturation, shot noise at the peak and additive baseline offsets.
2) AREA normalisation divides each spectrum by its integrated intensity over a chosen wavelength range. It is robust to multiplicative scaling but inherits sensitivity to the choice of integration window and cannot correct for additive baseline drift.

3) SNV (the standard normal variate transform, equivalent to z-score) is applied independently to each spectrum where SNV subtracts the per-spectrum mean and divides by the per-spectrum standard deviation following [38]:

$$x_{SNV,k} = \frac{x_k - \bar{x}}{s}, \bar{x} = \frac{1}{n}\sum_k x_k , s = \sqrt{\frac{1}{n-1}\sum_k (x_k - \bar{x})^2} \tag{6}$$

where $x_k = S_i(\lambda_k)$ is the intensity of a given spectrum $S_i$ at its $k^{th}$ wavelength channel (cropped to selected wavelengths, $k = 1,\ldots,n$), $\bar{x}$ and $s$ are, respectively, the mean and the standard deviation of the $n$ wavelength channels of that same spectrum, computed independently for every spectrum. SNV simultaneously removes additive baseline components and multiplicative intensity scaling, yielding a pure shape representation. For $Yb^{3+}$ thermometry, where the temperature information resides primarily in band-shape deformations rather than absolute intensities, SNV is the optimal choice and is therefore retained as the default for the benchmark presented in Section 4.

### 3.3 Principal Component Analysis: A Critical Step

Principal component analysis is a linear dimensionality reduction technique that transforms a set of possibly correlated variables into a set of linearly uncorrelated variables called principal components [39,40]. In the present context each spectrum $S_i$ is a point in a high-dimensional space with dimensionality equal to the number of spectral channels. PCA identifies the directions (eigenvectors) in this space along which the variance is maximised. Given a set of $N$ spectra $S_1, \ldots, S_N$, the procedure consists of mean-centring the data, computing the spectral covariance matrix and projecting each spectrum onto the first $d$ eigenvectors of that matrix:

$$[C]_{n\times n} = \frac{1}{N}\sum_i (S_i - \bar{S})_{n\times 1}(S_i - \bar{S})^T_{1\times n} , [PC_i]_d = (S_i - \bar{S})^T_{1\times n}[V_d]_{n\times d} \tag{7}$$

where $\bar{S}$ is the mean training spectrum, $V_d$ is the matrix whose columns are the first $d$ eigenvectors of $C$:

$$[C]\vec{v}_j = \lambda_j \vec{v}_j, j = 1, \ldots, n, [V_d]_{n\times d} = [\vec{v}_1 \quad \cdots \quad \vec{v}_d] \tag{8}$$

with eigenvectors sorted in decreasing order, $\lambda_1 \geq \ldots \geq \lambda_n$, so that $V_d$ retains the $d$ eigenvectors of largest variance. $PC_i$ is the $d$-dimensional score vector representing the $i$-th spectrum in the reduced space. The eigenvalues of $C$ encode the variance explained by each component; their rapid decay justifies dimensionality reduction from a few hundred spectral channels down to $d = 3$ components for the present dataset.

The principal-component loadings, i.e. the eigenvectors $v_j(\lambda)$ as functions of wavelength, admit a direct physical interpretation for $Yb^{3+}$ luminescence. PC1 captures the bulk temperature-induced redistribution of intensity between Stark sub-bands: as the temperature increases, the higher-energy Stark levels of $^2F_{5/2}$ become thermally populated and their emission intensities increase relative to the lowest Stark component. The PC1 loading therefore exhibits a characteristic bipolar pattern, with positive weights at long wavelengths (low-energy transitions) and negative weights at short wavelengths (high-energy transitions). PC2 and PC3 capture residual spectral asymmetries, fine structure and secondary broadening effects that PC1 cannot represent, including differential broadening of individual Stark transitions due to varying electron–phonon coupling strengths, asymmetric intensity shifts arising from crystal-field changes with thermal expansion, and subtle baseline curvature variations.

As already mentioned, prior work used only the first principal component as the thermometric parameter. While that univariate PC1 approach represented a significant advance over traditional LIR, it inherits a structural limitation: by discarding PC2, PC3 and higher components it ignores complementary information that, although small in variance contribution, is statistically significant for temperature prediction. Multi-component models (e.g. that exploit PC1+PC2+PC3) consistently outperform PC1-only regressors, for which a quantitative comparison on the $Yb^{3+}$ dataset is provided in Section 4. Conceptually, the higher PCs capture features such as differential broadening and crystal-field asymmetries that remain orthogonal to PC1 and therefore add a near-independent channel of temperature information. Single-component approaches should be viewed as a simplified, sub-optimal variant of the full LT 2.0 framework.

## 3.4 Machine-Learning Methods

Nineteen supervised regression algorithms drawn from four families with distinct mathematical foundations, inductive biases and suitability for luminescence thermometry are benchmarked in this work. The remainder of this section introduces each family, with equations provided where they convey the inductive bias most concisely. Quantitative performance figures for the $Yb^{3+}$ dataset are deferred to Section 4.

### 3.4.1 Tree-based ensemble methods

Decision-tree ensembles combine predictions from many decision trees to improve accuracy and robustness. Individual trees partition the input feature space into axis-aligned rectangular regions and predict a constant value within each region. Ensembles aggregate these piecewise-constant predictions through averaging, as in Random Forest [31], and Extra Trees [41], or through stage-wise additive boosting, as in Gradient Boosting, XGBoost [32], LightGBM [42], and CatBoost [43]. These methods are popular for tabular regression because of their flexibility and absence of feature-scaling requirements. They share, however, a structural property that becomes critical for continuous thermometry: the prediction at a query point is always a (possibly weighted) average of training labels and therefore lies strictly inside the convex hull of the training temperatures. The implications of this limitation are analysed in Section 4.

### 3.4.2 Physics-aware and smooth regression models

Physics-aware models exploit the a-priori knowledge that principal-component scores vary smoothly and often monotonically with temperature. Constraining the model to produce smooth predictions injects an inductive bias that improves generalisation and enables reliable interpolation. This family includes polynomial regression on PC1 with a Bayesian-information-criterion-selected degree, the monotone piecewise cubic Hermite interpolating polynomial (PCHIP) on PC1, which guarantees a strictly increasing or decreasing calibration curve [44], multivariate polynomial regression on (PC1, PC2, PC3), which captures interaction terms $PC_j \cdot PC_k$, the 3D spline on (PC1, PC2, PC3), and the inverse-variance sensor-fusion (SF) estimator.

The SF concept was introduced to luminescence thermometry by treating each temperature-dependent spectral feature as an independent sensor and combining the individual readings by inverse-variance weighting [15]. In that formulation, each channel $j$ is first inverted through its own calibration curve to yield a separate temperature estimate $\hat{T}_i = g_j^{-1}(PC_j)$, and the fused estimate is the precision-weighted average:

$$\hat{T} = \frac{\sum_j w_j \hat{T}_j}{\sum_j w_j}, w_j = \frac{1}{\sigma_{T,j}^2} \quad (9)$$

where $\sigma_{T,j}$ is the temperature uncertainty of channel $j$. This is optimal when every calibration curve is individually invertible (i.e. strictly monotonic) and locally linear, so that each channel can be turned into a stand-alone thermometer before averaging.

The estimator used here generalizes this idea rather than applying it directly. Each retained principal component is again treated as an independent sensor channel with a forward calibration curve $g_j(T)$, the mean score of component $j$ at temperature T, and a temperature-dependent uncertainty $\sigma_j(T)$ given as the standard deviation of that component across repeated spectra. Both $g_j(T)$ and $\sigma_j(T)$ are obtained by monotone (PCHIP) interpolation of the per-temperature means and standard deviations estimated on the training set. Instead of inverting each channel separately and then averaging the resulting temperatures, the temperature of a query spectrum $i$ is estimated by a single joint maximum likelihood inversion in PC space, selecting the temperature whose predicted coordinates best match all observed scores simultaneously:

$$\hat{T}_i = \arg\min_T \sum_{j=1}^{d} \frac{\left(PC_{j,i} - g_j(T)\right)^2}{\sigma_j^2(T)} \tag{10}$$

The minimizer is the maximum-likelihood temperature and weighting each channel by its precision $1/\sigma_j^2(T)$ automatically assigns greater influence to the components that are better resolved at a given temperature and down-weights those with greater uncertainty. This is a standard principle of metrological sensor combination, without tuning of any hyperparameter.

The practical advantages from solving the joint problem directly instead of averaging per-channel readings are loss of requirement of each $PC_j$ to have a monotonic, individually invertible calibration curve, and the use of the full temperature-dependent uncertainty at prediction time rather than a single scalar variance per channel, which matters when the spectral shape and its dispersion vary significantly across the measured range.

### 3.4.3 Kernel and instance-based methods

Kernel methods and instance-based learners make predictions based on local neighbourhoods in feature space and can capture complex, non-linear relationships without assuming a specific global functional form. The k-nearest-neighbours regressor (KNN) predicts temperature as the average of the k closest training labels in PCA space; like tree ensembles it suffers from a strict interpolation bound. Support-vector regression (SVR) constructs a function that ignores residuals smaller than a tolerance ε and penalizes only the excess beyond this ε-insensitive margin, while maintaining smoothness through regularization [45]. Both the radial-basis-function and polynomial kernel variants are evaluated here. Gaussian-process regression (GPR) is a Bayesian non-parametric method that places a prior distribution over functions and updates it with training data to obtain a posterior [46]. The posterior mean provides point predictions while the posterior variance provides calibrated uncertainty estimates that are valuable for deployment.

### 3.4.4 Linear-flavoured and neural-network models

This family includes polynomial kernel-ridge regression, partial least squares (PLS) on the spectral features, Bayesian ridge regression on polynomial features, ElasticNet regression with mixed $l_1/l_2$ regularization [47], and neural-network architectures. Two multilayer-perceptron (MLP) variants operate on (PC1, PC2, PC3) inputs: one trained with the Adam optimiser and hidden layers [128, 128, 64], the other trained with L-BFGS and hidden layers [256, 128, 64]. Both use rectified-linear-unit activations and weight-decay regularisation. A further MLP with hidden layers [256, 128, 64] is trained directly on the 124-channel SNV-normalised spectra without PCA. Finally, a shallow 1D convolutional neural network (CNN) with kernel size 7 is trained on the same raw SNV-normalised spectra. CNN are designed to exploit local spatial structure in sequential data through learnable filters [25].

## 4. Results and Discussion

### 4.1 The SpectraSensML Thermometry Framework: Open-Source Software

To facilitate the adoption of LT 2.0 methods across the luminescence-thermometry community, we developed SpectraSensML, an open-source application released under the GPLv3 license. The source code and installation files are available on GitHub (https://github.com/aleksandarciric83/SpectraSensML/releases/). The application implements the complete workflow from raw spectral data to trained models, thermometric predictions, and performance evaluation. It is distributed as self-contained installers for Windows, macOS, and Linux and does not require users to install Python or use command-line tools.

The application includes configurable baseline subtraction using user-defined background windows; spectral cropping to a selected emission range; MAX, AREA, and SNV normalization, LIR diagnostics with selectable integration windows, PCA analysis, and the nineteen regression algorithms described in Section 3.4. Model performance is reported using a unified set of metrics, including RMSE, MAE, bias, temperature resolution ($\delta T$), 95$^{th}$ percentile error, and maximum error, evaluated per split, per set point, and for the complete dataset. A single global random-number-generator seed controls all stochastic processes, ensuring bit-exact reproducibility.

The application was designed to be user friendly and compatible with multiple input data formats. Its workflow is described below (see Figure 3).

1. The user first loads the folder containing the spectral dataset. This folder should include subfolders corresponding to spectra recorded at different temperatures. Each subfolder should contain spectra measured at the same temperature.
2. The default training/validation split is 80:20; however, the user may freely adjust this ratio in the corresponding input field.
3. Although the application is configured by default for temperature sensing, it can also be used for other sensing applications, such as pressure or oxygen-concentration sensing [48–50]. The measured variable and its unit can be defined by the user.
4. Spectral files may include metadata, for example, information about the integration time or other acquisition parameters,
5. as well as information specifying the line at which the spectral data begin. When present, these details should be entered into the corresponding text fields in the application.
6. The user must also specify the extension of the spectral data files; the default extension is .txt.
7. Once settings 1 to 6 have been defined, clicking the “Load Spectra” button imports the spectral data into the application. The data are then automatically assigned a temperature and dataset role according to their subfolder names.
8. When subfolders follow the naming convention shown in Figure 3, the application automatically identifies both the temperature and the dataset role: training/validation or *test_unseen*. After loading, the user may still manually modify the temperature and role assigned to each folder. A prefix indicates whether the temperature is negative or positive, while the three-digit number following m or p defines the temperature in degrees Celsius. The value is subsequently converted to kelvin. By default, subfolders corresponding to temperatures divisible by 10 K are assigned to the training/validation dataset, whereas all remaining temperatures are assigned to the *test_unseen* dataset.
9. The selected folder assignments are confirmed by clicking the “Apply folder roles – dataset” button.

10. The user then proceeds to the “2 – Pre-process” tab.
11. Because only a selected portion of the recorded spectrum is typically relevant for the analysis, the lower and upper wavelength boundaries can be defined in dedicated text fields.
12. The user then selects the preprocessing method described in this work, with or without constant-background removal. SNV normalisation is selected by default.
13. For comparison with machine-learning approaches, the LIR section allows the user to define the lower and upper wavelength limits of the two emission bands used to calculate the intensity ratio.
14. After entering these settings, preprocessing is performed by clicking “Run Pre-Processing”.
15. The user then proceeds to the “3 – PCA” tab, where the complete PCA analysis is performed.
16. By default, three principal components are selected, however, the user may adjust this number.
17. Clicking “Run PCA” performs the PCA analysis and generates the associated graphical outputs.
18. The next tab is dedicated to machine-learning model selection.
19. The complete list of models described in this work is available and grouped according to model family.
20. After selecting the desired models, the user proceeds to the “Run” tab, where the machine-learning analysis is executed.
21. The user first selects an output directory in which all results will be generated automatically following a successful run.
22. The analysis is initiated by clicking “Run Benchmark”. Depending on the dataset size, selected models, and available hardware, the calculation may require several minutes. The CNN model is the most computationally demanding option and may be unsuitable for low-performance hardware.
23. The results are displayed in the subsequent tab.
24. Results are presented in a table ranked by thermometric performance, together with the corresponding evaluation metrics. The complete output can be exported as a CSV file by clicking the button below the table.
25. Generated figures can be selected and previewed in the designated section.
26. These figures can also be exported either as high-resolution, publication-ready raster images or in scalable vector graphics (SVG) format.

For the more detailed guide the user is referred to the end of the Supplementary Information document.

The output generated by SpectraSensML is used in the subsequent analysis of $Yb^{3+}$ thermometry. The corresponding dataset is provided together with the application, enabling full reproduction of the reported results. The software is provided free of charge under the condition that all research findings, analytical results, or figures generated using the application provide proper attribution by citing this paper.

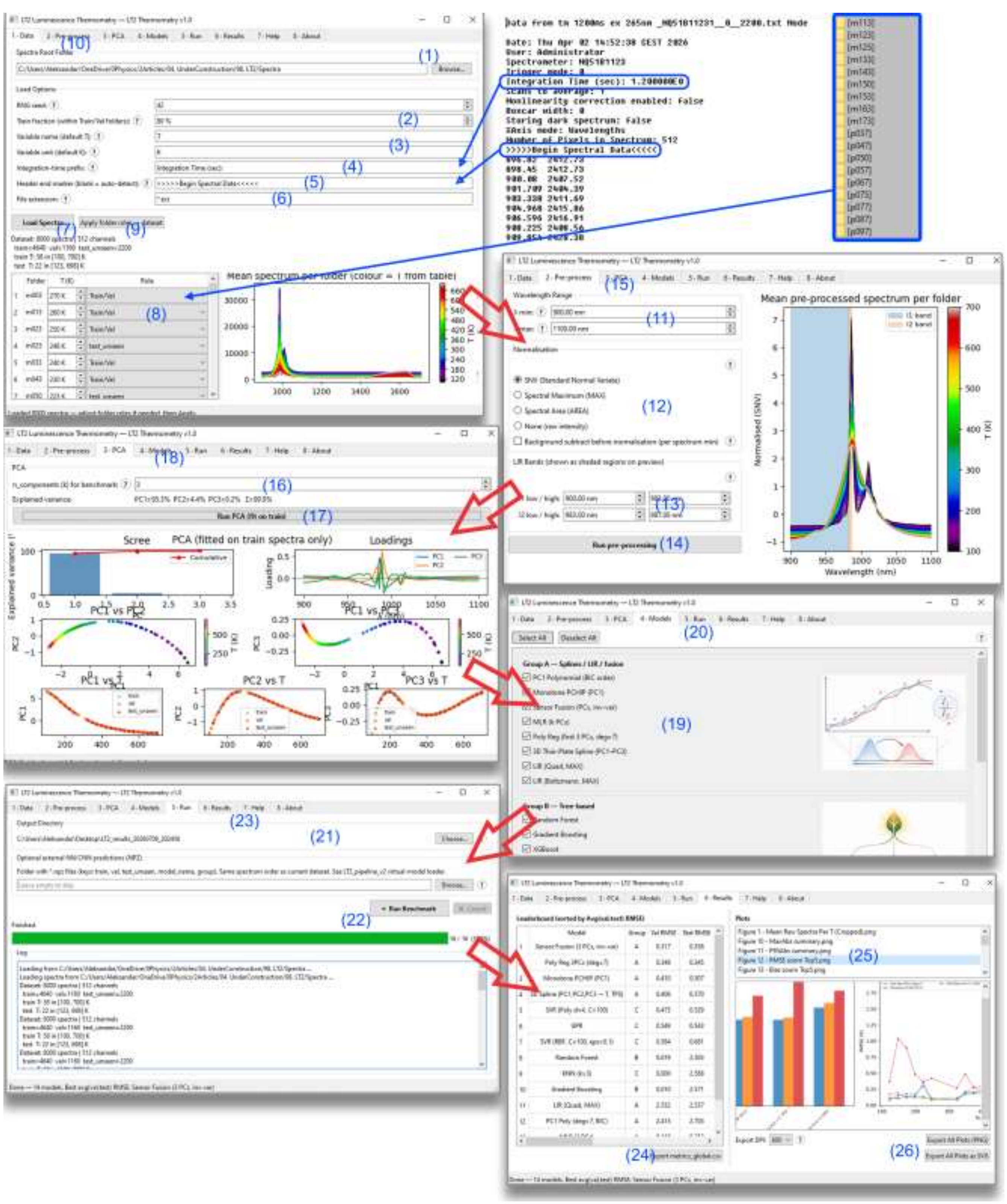

**Figure 3. SpectraSensML software application workflow.**

## 4.2 Spectral Characteristics and Temperature Dependence

Figure 4a shows the temperature evolution of the $Yb^{3+}$ emission band across 125 to 700 K. Two effects are visible by eye: a continuous redistribution of intensity between Stark sub-bands as the lattice expands and the phonon population grows, and a gradual line-shape broadening driven by electron-phonon coupling. Both effects are weak in any single wavelength channel, yet they are strong, smooth and monotonic when viewed as global band-shape deformations. This is the underlying motivation for full-spectrum machine learning.

Three normalisation strategies, MAX, AREA and SNV, were compared by training identical multilayer-perceptron and polynomial-regression pipelines on each. SNV normalisation consistently outperformed the alternatives across all model families, yielding 15 to 20 % lower RMSE on the validation set than AREA, and 25 to 40 % lower RMSE than MAX. The superior performance of SNV arises from its simultaneous removal of additive baseline offsets (dark current, stray light) and multiplicative intensity scaling (excitation-power fluctuations), leaving only the pure spectral shape, which is the feature that carries temperature information for $Yb^{3+}$. SNV is therefore adopted as the default normalisation for all subsequent results (see Figure 4b).

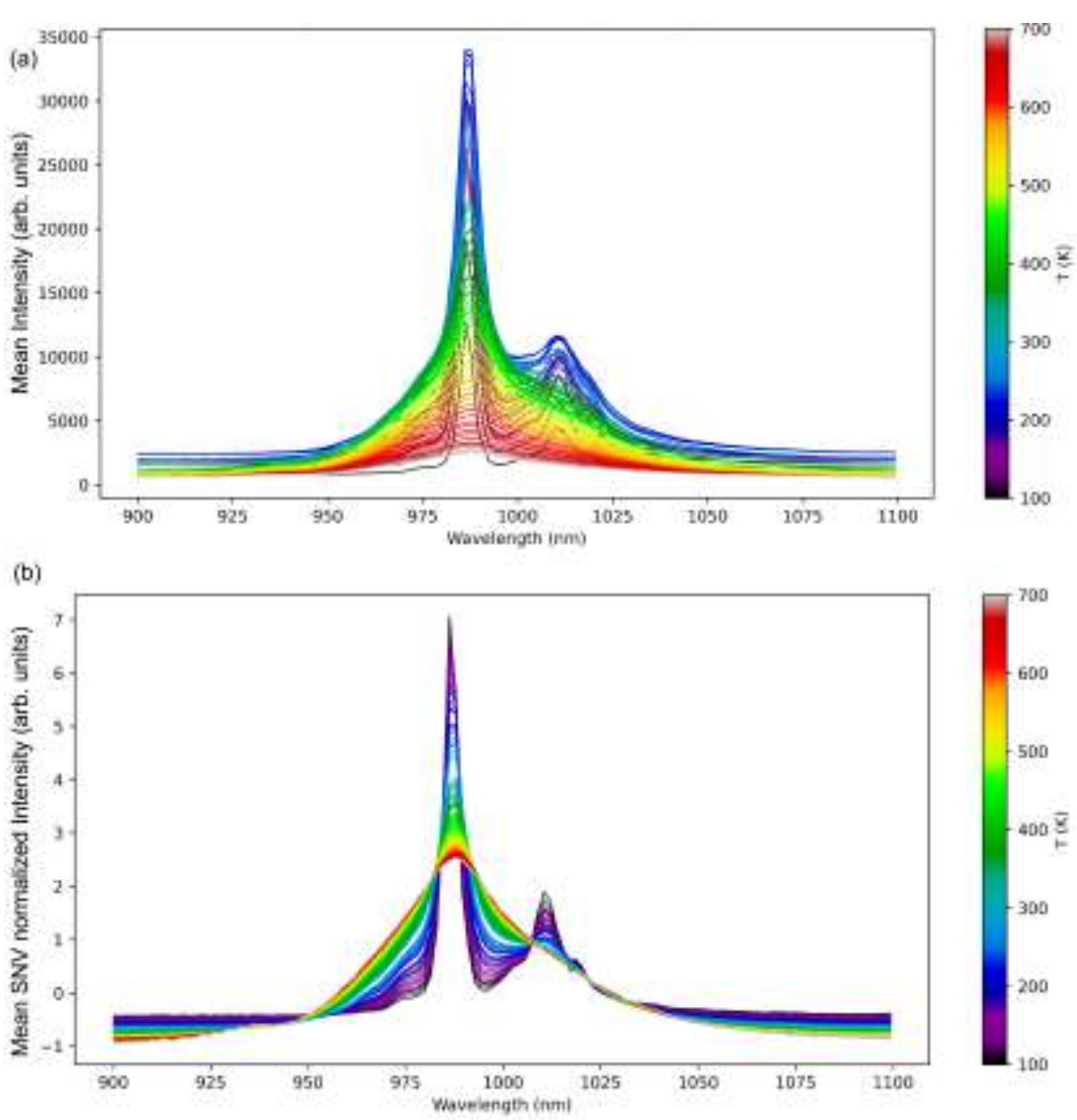

Figure 4. (a) Raw and (b) SNV-normalised $Yb^{3+}$ NIR emission spectra as a function of set-point temperature.

## 4.3 PCA Analysis Results

Principal-component analysis on SNV-normalised training spectra reveals that the first three components capture 99.94 % of the total variance, with PC1 accounting for 95.28 %, PC2 for 4.44 % and PC3 for 0.22 % Figure 5a. The rapid decay of explained variance justifies the reduction from 124 spectral channels to $d$ = 3 principal components. The PC1 loading exhibits the expected bipolar pattern that reflects temperature-induced Stark redistribution (Figure 5b). The PC1 score is a smooth, near-monotonic function of temperature over the entire 125 to 700 K range (also see Supplementary Information), well approximated by a degree-7 polynomial. PC2 and PC3 loadings show more complex, oscillatory patterns corresponding to differential broadening and fine-structure asymmetries, with scores exhibiting non-monotonic temperature dependence, capturing residual spectral features orthogonal to PC1. The smooth trajectories in (PC1, PC2, PC3; Figure 5c,d) space validate the use of smooth regression models. Conversely, any model that partitions the feature space into discrete regions, including trees and k-nearest-neighbours, will fail to capture the continuous temperature evolution.

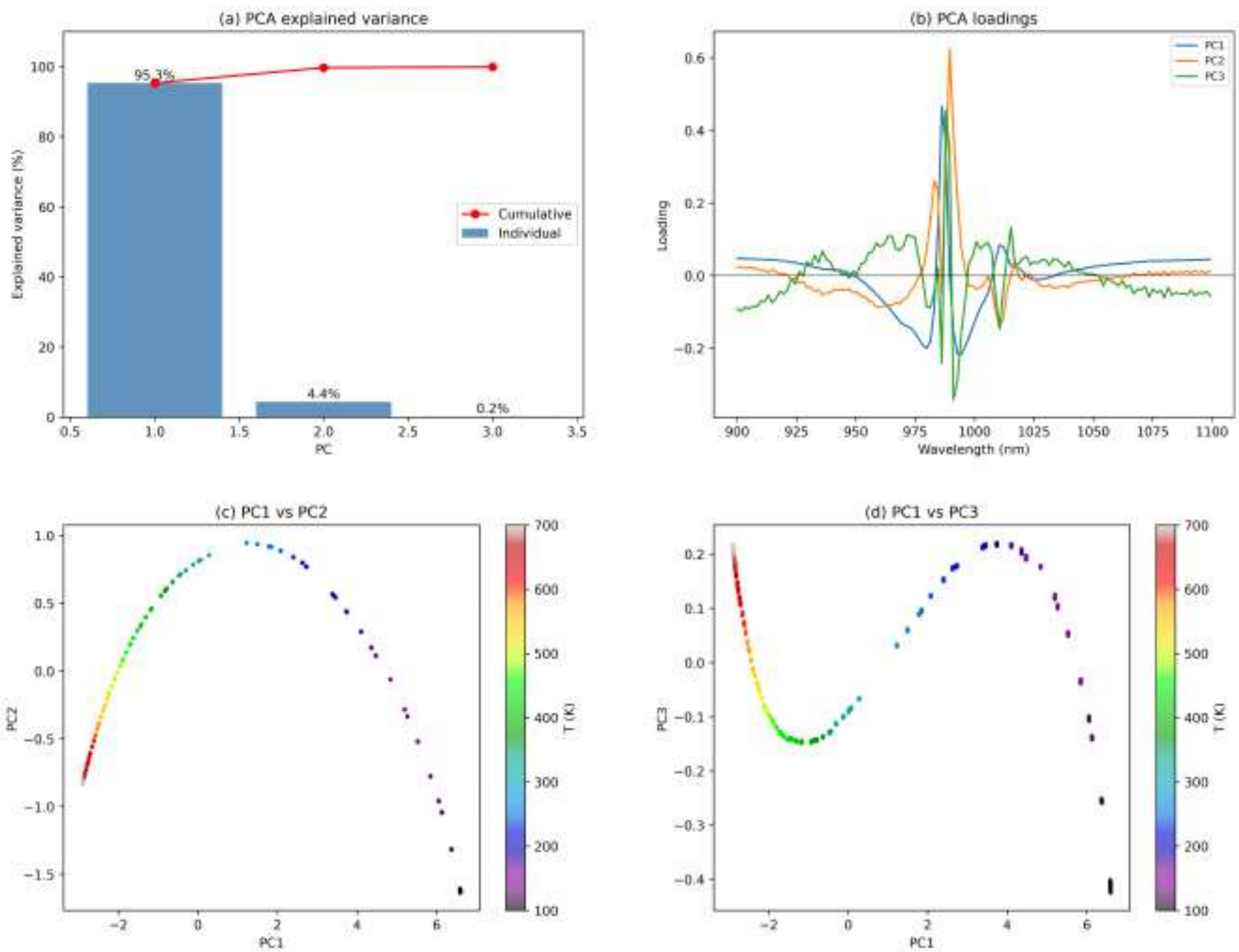

**Figure 5. PCA on SNV-normalised training spectra. (a) Scree plot showing cumulative explained variance versus component index. (b) Loadings of PC1–PC3 as a function of wavelength. (c, d) PC1–PC2 and PC1–PC3 score scatter plots colour-coded by temperature.**

## 4.4 Machine-Learning Benchmark

Supplementary information summarises the complete performance of all nineteen regression algorithms ranked by the mean of *validation* and *test_unseen* RMSE, while Figure 6 shows only the top 5 performers. The top of the leaderboard is occupied by smooth, multi-component regressors: the sensor-fusion estimator, the multivariate polynomial regression on (PC1, PC2, PC3) at degree 7, Extra Trees, the degree-4 polynomial kernel-ridge regressor, and the monotone PCHIP on PC1 alone. Monotone PCHIP on PC1 is more than five-fold better than the degree-7 polynomial on the same input, demonstrating the strong effect of injecting a monotonicity prior on a univariate calibration curve. For comparison with the traditional method, the quadratic LIR trained on MAX-normalised intensity ratios reaches RMSE about seven times worse than the best LT 2.0 model on identical data, and the Boltzmann LIR is unusable.

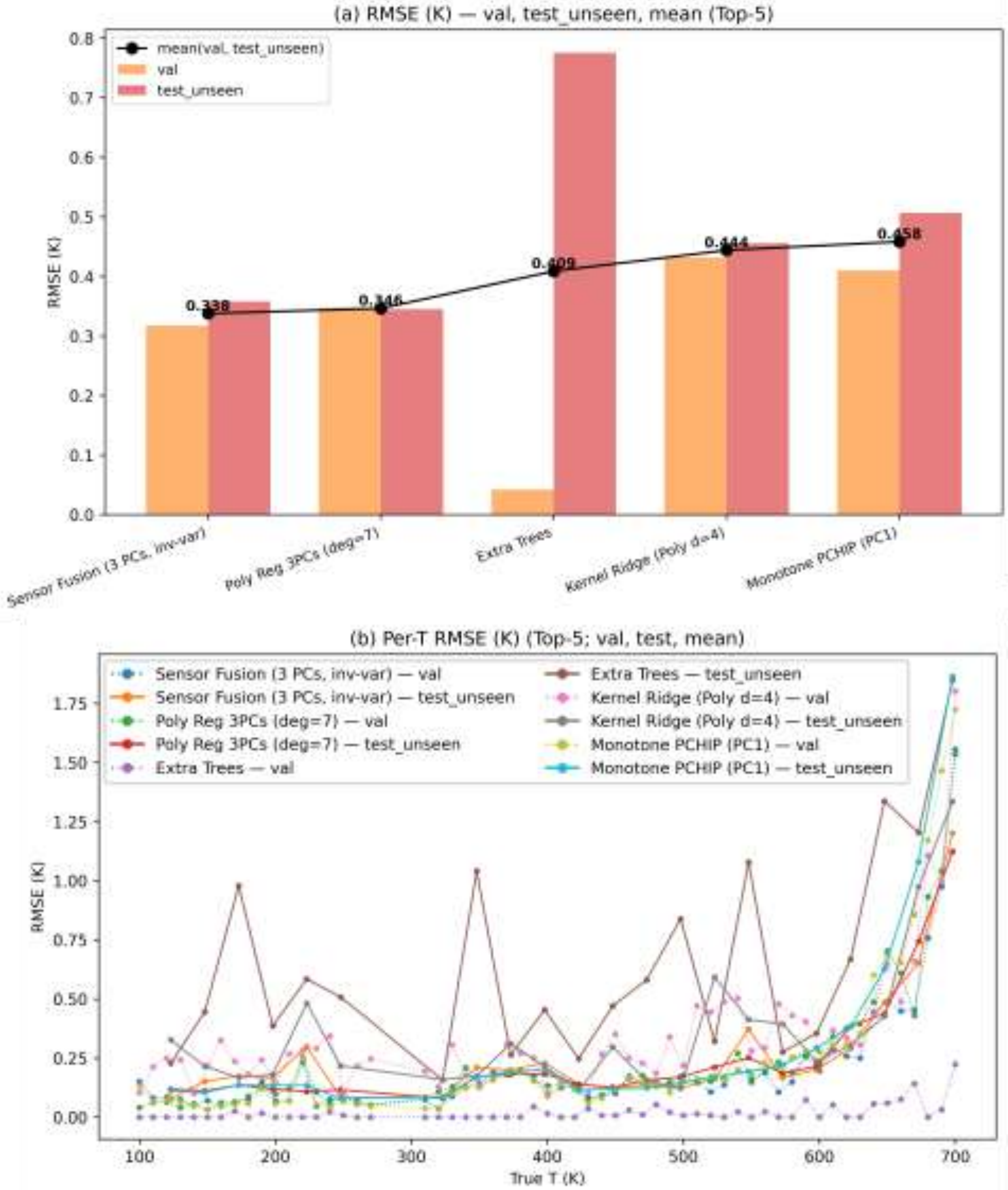

**Figure 6. Top performing models based on (a) mean of *validation* and *test_unseen* spectra, and (b) their per temperature RMSE.**

Tree-based ensembles exhibit a characteristic pattern: near-zero training and validation error, indicative of memorisation of the training set-points. Surprisingly, Extra Trees had relatively low uncertainty on unseen test set, the lowest of the tree-based models, which is what brought it to the top performing models. k-nearest-neighbours follows the same trend of this type of models having a high mean RMSE value. Across the leaderboard, the spread of *validation* RMSE is much smaller than the spread of *test_unseen* RMSE, confirming that the unseen-temperature split exposes generalisation behaviour that is invisible on an in-distribution validation set. Tree ensembles and KNN fail because they violate the smoothness requirement. Every tree partitions the input space into axis-aligned rectangles and returns a constant prediction within each rectangle, so a query spectrum that lies between two adjacent training set-points is mapped to one of them and never to the temperature in between. The resulting predicted-temperature curve is step-like, oscillating around the unseen set-points. Even XGBoost, LightGBM and CatBoost, which are widely used elsewhere in machine learning, inherit this convex-hull constraint and exhibit large systematic errors on the unseen-temperature split. KNN with k = 3 follows the same pattern at with high RMSE. Convolutional neural networks (CNN), while smooth in principle, are here penalised by the modest diversity of the dataset (~80 distinct temperatures) and by the high wavelength channel correlation that PCA already captures with three components.

The remaining four top models share two structural properties: (i) they produce smooth, low-degrees-of-freedom mappings that respect the smooth temperature trajectory through PCA space, and (ii) they make explicit use of multiple principal components, which lets them exploit complementary information missed by PC1 alone. The sensor-fusion estimator goes one step further by automatically weighting the individual PC-based predictors by the inverse of their per-temperature variance on the validation set, a metrology principle borrowed from instrumented sensor combination. The result is a model that consistently matches or exceeds the performance of its individual constituents while remaining fully interpretable and modular: each univariate regressor can be inspected, audited or replaced independently.

### 4.5 Comparison with Conventional LIR Thermometry

The conventional Luminescence Intensity Ratio (LIR) analysis employs two integration windows centered on the primary emission peak and the higher-energy region (Figure 7a). While the Boltzmann equation can model the resulting intensity ratio, its utility is restricted to high-temperature regimes (Figure 7b). In contrast, a quadratic fit aligns well across all temperatures and datasets, though it lacks physical significance (Figure 7c). Notably, the Boltzmann-based LIR fails to accurately calibrate to the true temperature over the full experimental range (Figure 7d), resulting in a mean RMSE of 2.54 K on the unseen test set for the quadratic approach (Figure 7e). This is due to the relatively high bias on the entire temperature range (Figure 7f). By comparison, the sensor-fusion LT 2.0 model achieves a significantly improved RMSE of 0.36 K on identical $Yb^{3+}$ data, which presents a sevenfold enhancement. This performance gain stems from the model's ability to extract information encoded in subtle intensity redistributions, differential broadening, and spectral asymmetries that two-window LIR ignores. Essentially, LT 2.0 functions as a generalized, data-driven LIR that processes all wavelength channels simultaneously, implicitly weighting them by information content. This approach overcomes the inherent limitations of feature-based LIR in $Yb^{3+}$ luminescence, where strongly overlapping peaks and a reliance on Stark sublevels hinder traditional calibration. Additional metrics for the LIR analysis are provided in the Supplementary Information.

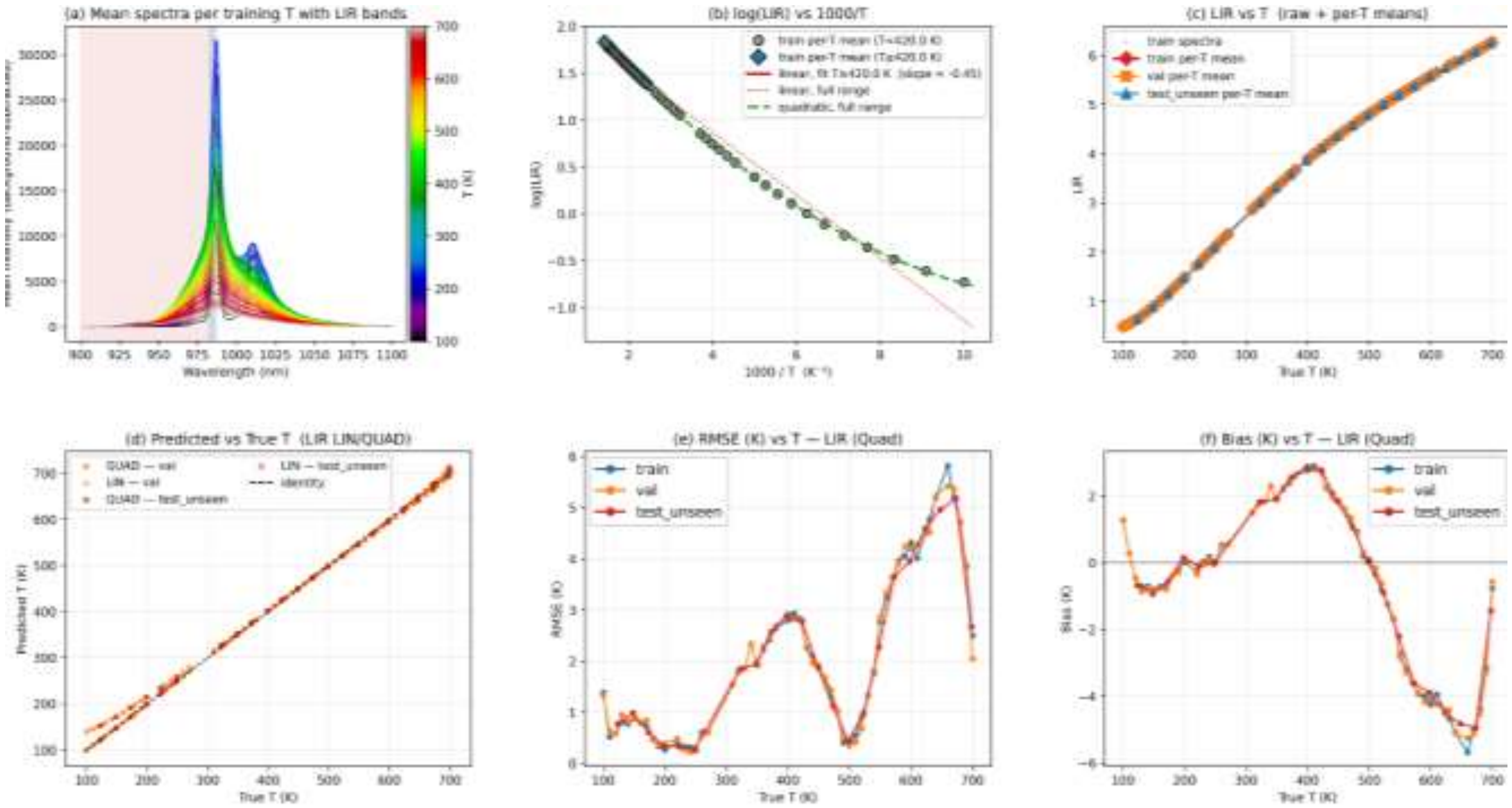

**Figure 7. (a) Mean (per temperature) spectra with marked spectral areas used for LIR. (b) LIR fitted to quadratic function and to the Boltzmann equation only at high temperatures, on train datasets. (c) LIR calibration on training set and use on validation and unseen data. (d) Predicted vs true temperature by using Boltzmann equation LIR and quadratic fits on validation and unseen datasets. (e) RMSE and (f) bias (accuracy) on quadratic fit of all three datasets.**

## 4.6 General Model-Selection Guidelines for LT 2.0

The $Yb^{3+}$ benchmark establishes results that should be interpreted with appropriate generality. Two classes of findings generalise robustly across emitters, but one does not.

First, these failure modes are general and do not depend on the specific luminescent emitter or spectral region. Tree-based ensemble methods and KNN regression cannot reliably interpolate beyond the range of temperatures represented in the training data, because their limitation is structural rather than dataset-specific. Consequently, they are not suitable for continuous-readout thermometers intended to predict temperatures that were not included during training. Nevertheless, these approaches can be used appropriately for fixed-state sensing, in which the operating temperatures coincide with the trained set points, or for discrete sensing applications, such as multi-bit discrete sensors, quantised-output sensors, or digital encoders. To ensure broad applicability and provide users with a comprehensive selection of predictive methods, SpectraSensML includes both tree-based ensemble models and KNN regression. Their performance in continuous sensing should, however, be interpreted cautiously: high accuracy on interpolated training/validation data does not necessarily demonstrate reliable prediction at previously unseen temperatures. Evaluation using a genuinely unseen test dataset is therefore essential.

Second, the most promising families generalise. Smooth physics-aware regressors operating on a few principal components, kernel methods with smooth kernels (polynomial kernel ridge, SVR with RBF or polynomial kernel, Gaussian-process regression) and modestly regularised multilayer perceptrons on PCA features have all demonstrated competitive performance here and are expected to remain competitive on other lanthanide and transition-metal thermometers with similar smooth spectral evolution.

Third, the identity of the single best model does not generalise. The exact ranking of the top family members depends on the smoothness of the chosen activator's spectral evolution, the size of the dataset, the level of experimental noise and the temperature range. The sensor-fusion estimator is consistently competitive because of its modular structure, but for any new material the practical recommendation is to evaluate several models from the proven families on the same dataset and to rank them by the same protocol used here (mean of validation and *test_unseen* RMSE under a fixed splitting seed). The SpectraSensML Thermometry application described in Section 4.1 automates that procedure.

Thus, although the general guidelines presented in this paper hold for all the cases of continuous sensing by spectroscopy, the ranking of the model performance will depend on multiple factors from luminescent material to the experimental setup. It is necessary to test as many models as possible, for which the robustness and consistency of the SpectraSensML software prove indispensable.

## 5. Conclusions

A paradigm shift in luminescence thermometry has been demonstrated: spectrum-wise machine learning provides a substantial improvement over the conventional luminescence intensity ratio for $Yb^{3+}$ luminescence thermometry in the near-infrared biological transparency window. By transitioning from the traditional LT1.0 paradigm to a full-spectrum, data-driven approach, this study demonstrates a profound enhancement in thermometric performance. Specifically, for the demanding case of $Yb^{3+}$, the application of machine learning regression yielded an unseen-temperature RMSE of 0.36 K across the 125–700 K range. This represents a seven-fold gain in predictive performance over the best classical luminescence intensity ratio (LIR) variants evaluated on the exact same dataset, proving that comprehensive spectral analytics can overcome the inherent limitations of isolated feature tracking.

Crucial to facilitating this methodological shift is the introduction of SpectraSensML. This dedicated software platform provides the research community with an accessible, open-source framework to rigorously evaluate and deploy complex machine learning models without necessitating advanced programming expertise. By streamlining the workflow from raw spectral data to predictive temperature readouts, SpectraSensML bridges the gap between theoretical data science and practical optical metrology.

The standard normal variate transform is identified as the most effective preprocessing strategy because it isolates the band-shape changes that encode temperature. Principal-component analysis on SNV-normalised spectra captures 99.94 % of the total variance in just three components, and physics-aware regressors operating on these components, in particular multivariate polynomial regression, inverse-variance sensor fusion and L-BFGS-trained multilayer perceptrons, achieve sub-Kelvin mean RMSE on a fully held-out unseen-temperature test set, a seven-fold improvement over the best LIR variant on the same data.

The structural reason behind the failure of decision-tree ensembles on unseen temperatures has been explained: their piecewise-constant predictions cannot interpolate beyond the convex hull of training set-points and produce step-function calibration curves with systematic errors of several Kelvin. Tree ensembles should therefore be avoided for continuous thermometry and reserved for fixed-state sensing. Smooth regressors (polynomials, splines, kernel methods, Gaussian processes and modestly regularised neural networks) that respect the continuous temperature trajectory through spectral feature space should be preferred.

The LT 2.0 framework is not limited to $Yb^{3+}$ thermometry. The methodology generalises to any luminescent material whose emission spectrum encodes temperature information, including lanthanide-doped phosphors based on $Er^{3+}$, $Nd^{3+}$, $Tm^{3+}$, $Dy^{3+}$, $Eu^{3+}$, transition-metal ions such as $Cr^{3+}$ or $Mn^{4+}$, quantum dots, organic dyes and upconversion nanoparticles. Beyond thermometry, the same preprocessing and regression pipelines apply to any scalar physical quantity encoded in a structured spectrum. Future work should focus on community-wide benchmark datasets for the most studied emitters, transfer learning between materials to reduce the calibration-data burden, and physics-informed neural networks that incorporate Boltzmann priors and known energy-level structure as inductive biases. The open-source SpectraSensML framework released with this work ensures that all these directions can be pursued in a reproducible and methodologically consistent way, while future versions will include even more models. The SpectraSensML application (source code, documentation, pre-built installers for Windows, macOS and Linux, and example datasets) is publicly available from the GitHub (https://github.com/aleksandarciric83/SpectraSensML/releases/) and the dataset used in this paper is available at Zenodo (https://doi.org/10.5281/zenodo.21318397). The full $YVO_4:Yb^{3+}$ benchmark dataset are archived alongside the software to enable exact reproduction of all results reported here.

Ultimately, while this study showcases the Luminescence Thermometry 2.0 methodology using a specific near-infrared emitting phosphor, the data-driven framework and the SpectraSensML platform are fundamentally material-independent. This approach is poised to unleash the potential of comprehensive spectral analytics for unmatched precision across any luminescent sensor, shifting the field's focus toward robust, holistic predictive performance.

## Acknowledgements

The authors from Vinča Institute would like to acknowledge funding of the Ministry of Science, Technological Development, and Innovation of the Republic of Serbia under contract 451-03-33/2026-03/200017.